\documentclass[preprint,sort&compress,3p]{elsarticle}

\usepackage{amsmath,amssymb}
\usepackage{xspace}
\usepackage{longtable}

\usepackage{ulem}
\usepackage{color}
\usepackage[mathscr]{eucal}
\usepackage{pifont}
\usepackage{listings}
\usepackage{slashed}
\usepackage{cancel}
\usepackage{epstopdf}
\usepackage{url}
\usepackage{xcolor}

\definecolor{mygray1}{rgb}{0.3,0.3,0.3}
\definecolor{mygray}{rgb}{0.9,0.9,0.9}

\title{{\tt LOOL}: Mathematica package for evaluating\\ leading order one loop functions}

\author{Amon Ilakovac, Luka Popov}
\ead{lpopov@phy.hr}
\address{University of Zagreb, Department of Physics, Bijeni\v cka cesta 32, Zagreb}

\begin{document}
\begin{flushright}
{\tt ZTF-EP-14-08}
\end{flushright}

\begin{abstract}

One-loop functions with loop masses larger than external masses and momenta can
always be expanded in terms of external masses and momenta. The precision requested
for observables determines the number of the expansion terms retained in the evaluation.
The evaluation of these expansion terms turns out to be much simpler than the 
exact evaluation of the corresponding one-loop function. Here we present the 
program which evaluates those expansion terms.

This {\it Mathematica} package provides two subroutines. First one performs analytical 
evaluation of basic one loop integrals. The second one is used to construct
composite functions out of those integrals. Composite functions thus obtained are
ready for numerical evaluation with literary no time consumption.
\end{abstract}


\maketitle
\section{Introduction} \label{}

Evaluation of loop integrals is always a challenging technicality, 
both analytically and numerically. Consequently, there have been numerous 
attempts to make computer algorithms which will make these computations
both fast and automatic. Among various attempts, one can stress out
{\it Mathematica} packages such as \texttt{LoopTools} \cite{LoopTools},
and \texttt{ANT} \cite{Angel2013}.

Although these and similar packages can be extremely useful in numerous
calculations, they didn't quite match our needs when dealing with
charge lepton flavor violation (CLFV) processes \cite{Popov2013,Popov2013a}.
For that purpose, we have developed a package specialized in analytical and
numerical evaluation of the loop functions expanded with respect to the
momenta and masses of the external charged leptons, while keeping only
the leading non-zero terms. Some of these methods were discussed in the
Appendix B in Ref.~\cite{PopovPhD}, while the very concept of mass and momenta
expansion was discussed in 
Refs.~\cite{Veltman1977,Fleischer1994,Tarasov1995,Tarasov1996,Smirnov2002}.

This package is written for {\it Wolfram Mathematica}, with the idea of
being easy and straightforward to install and apply for various cases.

\section{Installation and startup}

The package can be downloaded from \url{http://lool.hepforge.org}. Once downloaded,
it can be loaded into the {\it Mathematica} notebook in a usual manner,

\begin{quote}
\fcolorbox{mygray}{mygray}{
\texttt{<< "/path/to/file/LOOL.m"}}
\end{quote}

\section{Basic integrals}

Any given one-loop amplitude can be expanded over the external masses and
momenta \cite{VanTran1984}. The expansion terms can in turn be expressed via dimensionless 
loop integrals.
\begin{align} \label{integrals}
\bar{J}^m_{n_1n_2\dots n_k} (\lambda_1,\lambda_2,\dots,\lambda_k)
&=
\frac{(\mu^2)^{2-D/2}}{(M^2_W)^{D/2+m-\sum_i n_i}}
 \int\frac{d^D\ell}{(2\pi)^D} \frac{(\ell^2)^m}{\prod_{i=1}^k (\ell^2
   - m_i^2)^{n_i}}
\nonumber\\
&= \frac{i(-1)^{m-\sum_i n_i}}{(4\pi)^{D/2}\Gamma(\frac{D}{2})}
 \Big(\frac{\mu^2}{M_W^2}\Big)^{2-D/2}
 \int_0^\infty \frac{dx x^{D/2-1+m}}{\prod_{i=1}^k (x+\lambda_i)^{n_i}} \,,
\end{align} 
where $m_i$ are  loop particle masses, $n_i$ are  the exponents of the
propagator  denominators,  $\lambda_i=m_i^2/M_W^2$  are  dimensionless
mass parameters  and $\mu$ is  't~Hooft's renormalization  mass scale.
For convergent integrals, one may set $D=4$, whilst for  divergent integrals one 
takes $D=4-2\epsilon$. Factor  $i/(4\pi)^2$ is pulled out from  all  integrals. 
Thus, for  finite integrals one obtains:
\begin{equation}
\bar{J}^m_{n_1n_2\dots n_k} (\lambda_1,\lambda_2,\dots,\lambda_k)
\ \equiv\
\frac{i}{(4\pi)^2}\ J^m_{n_1n_2\dots n_k} (\lambda_1,\lambda_2,\dots,\lambda_k)\; .
\end{equation}

On the other hand,  the divergent  integrals  are written  down  as a  sum of  a
divergent and constant term and a finite mass-dependent term:
\begin{equation}
\bar{J}^m_{n_1n_2\dots n_k}
(\lambda_1,\lambda_2,\dots\lambda_k)\ \equiv\
 \frac{i}{(4\pi)^2}\ \big[ \frac{1}{\varepsilon} + \mbox{const} +
 J^m_{n_1n_2\dots n_k} (\lambda_1,\lambda_2,\dots,\lambda_k) \big]\,.
\end{equation}
Due to GIM-like mechanisms (see for example Eqs.~(2.9) and (2.10) in
Ref.~\cite{Ilakovac1995}), the ``divergent+constant'' terms usually 
don't have a role in evaluating the amplitudes. In that case, all
amplitudes  can be  expressed in terms  of finite  mass dependent
functions $J^m_{n_1n_2\dots}  (\lambda_1,\lambda_2,\dots)$, which we
call the {\it basic loop integrals}.

These integrals can be evaluated using \verb+JInt+ subroutine, which
has the following syntax:
\begin{quote}
\fcolorbox{mygray}{mygray}
{\texttt{JInt[\{m, n1, n2, ...\}, \{x, y, ...\}]}}
\end{quote}
Notice that this function is called with two variable lists. 
First one stands for  $m, n_1, n_2, \ldots, n_k$, while the 
second one stands for $\lambda_1, \lambda_2, \ldots, \lambda_k$, 
as noted in Eq.~\eqref{integrals}. The parameter $k$ can assume
values $k = 1, 2, 3, 4$, corresponding to the tadpole, self energy,
triangle and box loop integrals, respectively. This can be easy
extended to any $k$-point function.

This function returns two expressions. First value correspond to the
expression for the basic loop integral $J^m_{n_1n_2\dots}  (\lambda_1,\lambda_2,\dots)$,
while the second one gives the accompanied ``divergent+constant'' terms.

\paragraph{Example 1} Evaluate the integral $J^1_{111} (x,y,z)$. 

\paragraph{Solution} According to the subroutine syntax, one easily obtains:
\begin{quote}
\begin{verbatim}
In[1]  := JInt[{1, 1, 1, 1}, {x, y, z}]
Out[1] := {(x^2*(-y+z)*Log[x] + y^2*(x-z)*Log[y] + 
          (-x+y)*z^2*Log[z])/((x-y)*(x-z)*(y-z)), 
          1 - EulerGamma + \[Epsilon]^(-1) + Log[4*Pi] + 2*Log[\[Mu]]}
\end{verbatim}
\end{quote}
Or written in the more readable fashion,
\begin{equation*}
 J^1_{111} (x, y, z) = \frac{x^2 (z-y) \ln(x) + y^2 (x-z) \ln(y) + z^2 (y-x) \ln(z)}{(x-y)(x-z)(y-z)} 
  + \frac{1}{\varepsilon} + 2 \ln(\mu) - \gamma + 1 + \ln(4\pi) \,,
\end{equation*}
where $\gamma=0.5772\ldots$ is the Euler-Mascheroni constant.

\section{Composite functions}

\begin{figure}[h!t]
  \centering
  \includegraphics[scale=0.6]{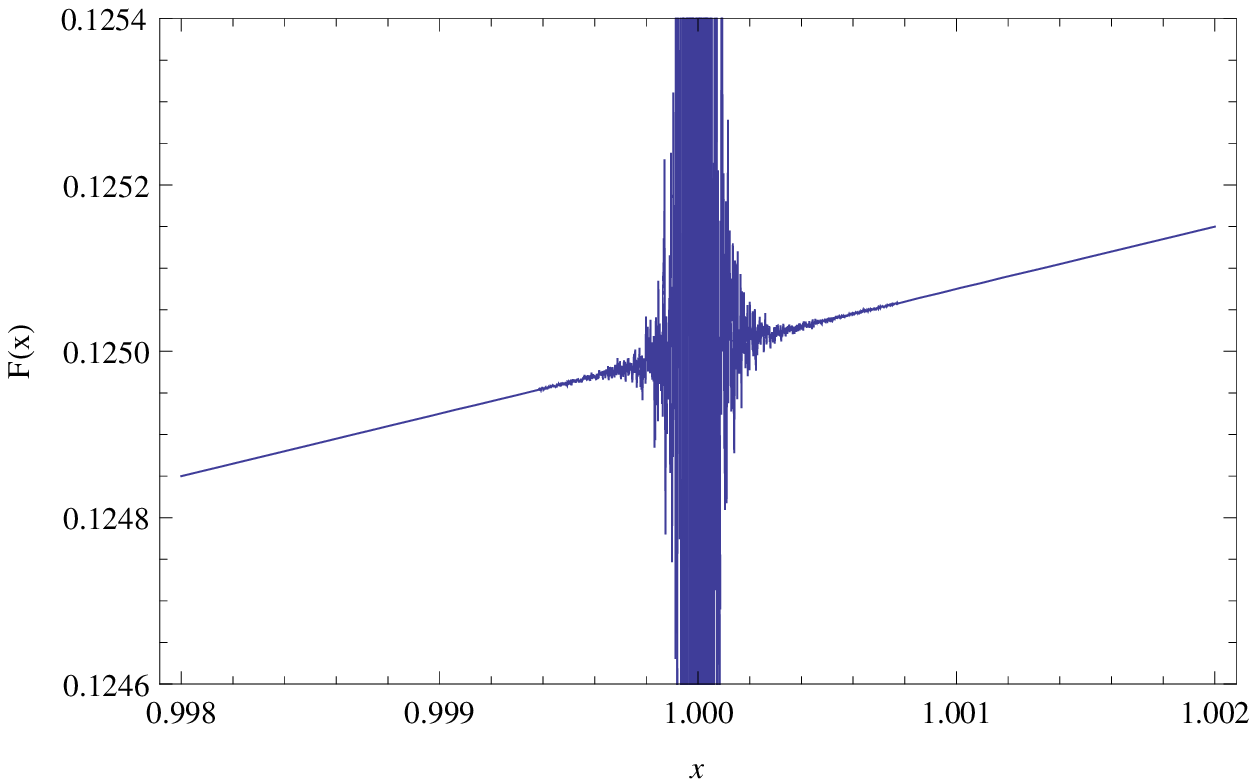} $\qquad$
  \includegraphics[scale=0.6]{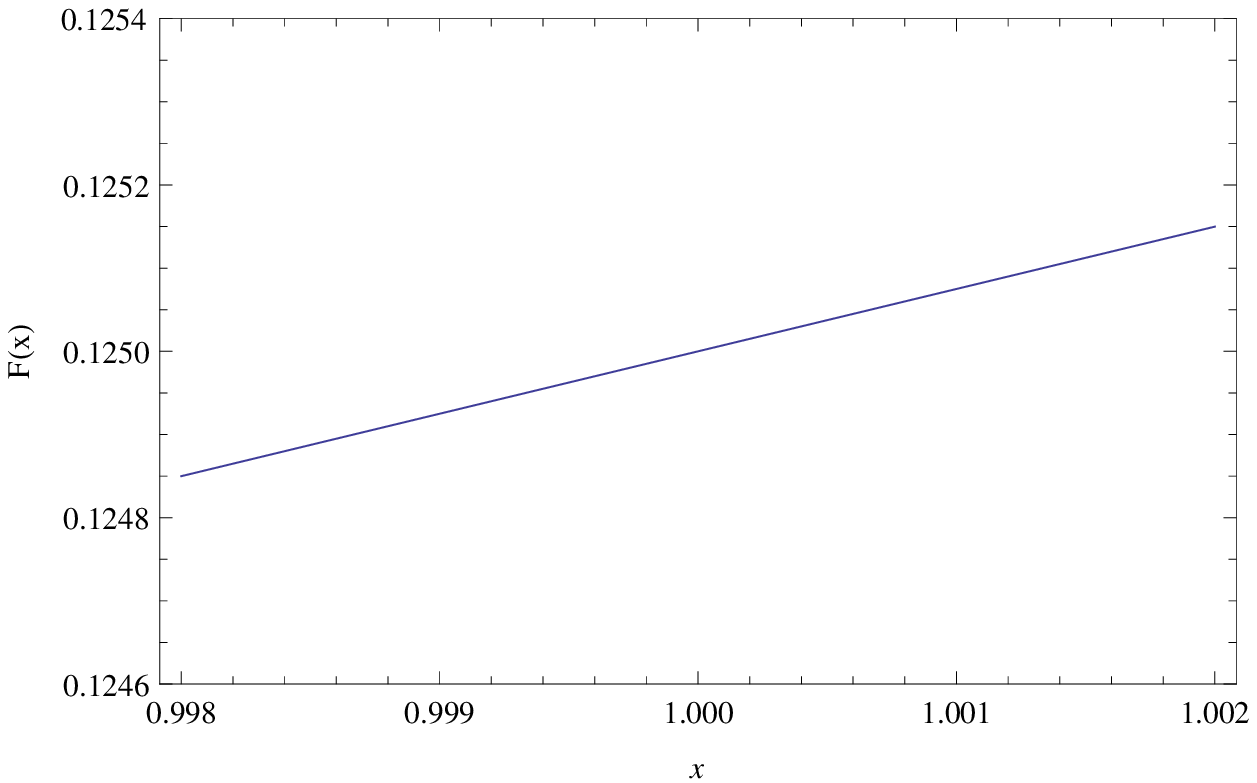}
\caption{Behavior of function 
$F(x) = J^1_{31}(1,x) + J^2_{41}(1,x) + x \left(
\frac{1}{2} J^1_{41}(1,x)-J^0_{31}(1,x)\right)+\frac{5}{6}$.
On the left panel one can notice the numerical instability for
critical values when
real numbers are used, while the right panel displays the evaluation of
the same function using rational numbers (obtained by the 
\texttt{CompositeF} function).} \label{fig1}
\end{figure}

For concrete applications, one wants to numerically evaluate functions which are
composed out of the $J$-integrals listed above. This can be done using the 
second subroutine, \verb+CompositeF+. This subroutine has the following
syntax:
\begin{quote}
\fcolorbox{mygray}{mygray}{
{\texttt{ CompositeF[fun, \{x, y, ...\}]}}}
\end{quote}
This function is called with two arguments. First argument stands for
some function composed out of $J$-integrals, and second one lists the
arguments of this very function. 

\verb+CompositeF+ subroutine results with function which is numerically
ready for evaluation, with all possible limits evaluated at once. The 
problem with numerical instability near the critical values
which often occurs in {\it Mathematica}
is elegantly solved by transforming decimal numbers into rational numbers
just before the evaluation. For this reason, the final numerical result 
will always be given as a rational number, which, if necessary, can be
expressed as a real number using {\it Mathematica} function \verb+N+.
The result of this approach can be seen in Fig.~\ref{fig1}.

The following example should further illustrate the use of \verb+CompositeF+.

\paragraph{Example 2} Construct the following composite function
$$F(x,y,z) = J^1_{1111}(1,z,x,y) + x y J^0_{1111}(1,z,x,y) + 
 \frac{x y}{\tan^2\beta} J^0_{211}(z, x, y) \,,$$
and numerically evaluate $F(0,1,1)$, supposing the existence of GIM-mechanism.

\paragraph{Solution} First step is to evaluate the basic loop integrals. Due to the 
assumed existence of GIM-mechanism, we are not interested in the constant terms. After
the definition of the relevant $J$-integrals, one can easily define the 
composite function and simply evaluate it for the given values.

\begin{quote}
\begin{verbatim}
In[1]  := J11111[x_, y_, z_, w_] = JInt[{1, 1, 1, 1, 1}, {x, y, z, w}][[1]]
Out[1] := -((w^2*Log[w])/((w-x)*(w-y)*(w-z))) + (x^2*Log[x])/((w-x)*(x-y)*(x-z)) + 
          (y^2*Log[y])/((w-y)*(-x+y)*(y-z)) + (z^2*Log[z])/((w-z)*(-x+z)*(-y+z))

In[2]  := J01111[x_, y_, z_, w_] = JInt[{0, 1, 1, 1, 1}, {x, y, z, w}][[1]]
Out[2] := -((w*Log[w])/((w-x)*(w-y)*(w-z))) + (x*Log[x])/((w-x)*(x-y)*(x-z)) + 
          (y*Log[y])/((w-y)*(-x+y)*(y-z)) + (z*Log[z])/((w-z)*(-x+z)*(-y+z))

In[3]  := J0211[x_, y_, z_] = JInt[{0, 2, 1, 1}, {x, y, z}][[1]]
Out[3] := ((y-z)*(x^2-y*z)*Log[x] + (x-z)*(-((x-y)*(y-z)) + y*(-x+z)*Log[y]) + 
          (x-y)^2*z*Log[z])/((x-y)^2*(x-z)^2*(y-z))

In[4]  := F[x_, y_, z_] = CompositeF[J11111[1, z, x, y] + x*y*J01111[1, z, x, y] + 
          (x*y)/tb^2*J0211[z, x, y], {x, y, z}];

In[5]  := F[0, 1, 1]
Out[5] := -1/2
\end{verbatim}
\end{quote}

While the first four steps do take certain amount of time to evaluate, last step
literary takes zero time to complete. That is the reason why it is of great importance
to use the equal sign ``\verb|=|'' rather than the definition sign ``\verb|:=|'' when defining
the functions. 

In order to make future evaluations time effective, one is advised to save the once 
evaluated composite function into a separate file,
%
%
\begin{quote}
\fcolorbox{mygray}{mygray}{
{\tt Save["some\_file.dat", \{F\}]}}
\end{quote}
which can be invoked in any other {\it Mathematica} notebook, for example
\begin{quote}
\begin{verbatim}
In[1]  := <<"\path\to\file\some_file.dat";

In[2]  := F[0, 1, 1]
Out[2] := -1/2
\end{verbatim}
\end{quote}

\section{Conclusion}

We have developed and presented a {\it Mathematica} package \texttt{LOOL} 
which calculates leading order one loop functions, both analytically and
numerically. The composite functions composed out of basic loop integrals are
evaluated only once, including all possible limits. Additionally, the problem
with numerical instability near the critical values is successfully  solved by 
dealing with rational rather to real numbers. 

\section*{Acknowledgments}

We thank Jiangyang You and Goran Popara for most useful discussions. 
This work has been fully supported by by the Croatian Science 
Foundation under the project MIAU.

\section*{References}
\bibliographystyle{elsarticle-num}
\bibliography{/home/lpopov/BibTeX/HEP_bibtex}

\end{document}